\documentclass[%
 amsmath,amssymb,
 aps,
 prb,
 twocolumn
]{revtex4-2}

\usepackage{graphicx}
\usepackage{dcolumn}
\usepackage{bm}
\usepackage{xcolor}
\usepackage[caption=false]{subfig}
\usepackage{tikz}
\usepackage{xfrac}
\usepackage{blindtext}
\usepackage{times}

\definecolor{darkblue}{HTML}{004D6B}
\definecolor{darkred}{HTML}{8c1515}

\usepackage{hyperref}
\hypersetup{
	colorlinks=true,
	urlcolor=darkblue,
	citecolor=darkred,
	linkcolor=darkblue,
	breaklinks
}

\begin{document}

\title{Recipes for the Digital Quantum Simulation of Lattice Spin Systems}

\author{Guido Burkard}
\affiliation{Department of Physics, University of Konstanz, D-78457 Konstanz, Germany}


\begin{abstract}
We describe methods to construct digital quantum
simulation algorithms for quantum spin systems on a regular lattice
with local interactions.  In addition to tools such as the Trotter-Suzuki
expansion and graph coloring, we also discuss the efficiency gained
by parallel execution of an extensive number of commuting terms.
We provide resource estimates and quantum circuit elements for the
most important cases and classes of spin systems.  As resource
estimates we indicate the total number of gates $N$ and simulation
time $T$,
expressed in terms of the number $n$ of spin 1/2 lattice sites (qubits),
target accuracy $\epsilon$, and simulated time $t$.  We provide
circuit constructions that realize the simulation time
$T^{(1)}\propto nt^2/\epsilon$ and $T^{(2q)}\propto
t^{1+\eta}n^\eta/\epsilon^\eta$
for arbitrarily small $\eta = 1/2q$ for the first-order and higher-order
Trotter expansions.  We also discuss
the potential impact of scaled gates, which have not yet been fully explored.
\end{abstract}

\maketitle

\section{Introduction}

The idea that the simulation of quantum systems, while
exponentially hard on a classical computer, can be
done efficiently on a quantum computer dates back
three decades, to Richard Feynman \cite{Feynman1982}.
A specific task of a digital quantum simulator consists
of \textit{Hamiltonian simulation}, i.e. the propagation
of an initial state $\psi(0)$ to the state
$\psi(t)$ at a later time $t>0$ according to
the dynamics generated by a Hamiltonian $H(\tau)$ via the
time-dependent Schr\"odinger equation,
$i\hbar \partial_\tau \psi(\tau)  = H(\tau) \psi(\tau)$,
for $0\le \tau\le t$.
Quantum simulation has been studied for a variety of
quantum systems on several simulator platforms \cite{Georgescu2014}.
Systems that can be simulated include quantum field theories \cite{Jordan2012},
quantum chemistry \cite{Wecker2014},
and fermionic lattice models \cite{Abrams1997,Wecker2015}.
Experimental demonstrations of quantum simulation
have been realized using superconducting circuits \cite{Barends2015,Lamata2018},
ion traps \cite{Monroe2021}, and semiconductor spin qubits
\cite{Xue2022}.

The resources required to simulate the time evolution of
a discrete quantum system on a quantum computer can be quantified in terms of the
system size (measured in the number of qubits $n$ required to store
its quantum state), as well as the duration $t$ and desired accuracy
$\epsilon$ of the simulation.
For a lattice of spins 1/2, $n$ directly represents the number of
lattice sites.
The measure we use to quantify
the simulation complexity is the duration $T=T(n,t,\epsilon)$
of the simulation, not to be confused with the simulated time $t$.

One of the early results of quantum simulation was the insight that quantum systems with local interactions can
be simulated efficiently on a quantum computer
using the Trotter decomposition \cite{Lloyd1996}.
The number $m$  of elementary, discrete time increments needed to
simulate the quantum evolution of a system during time $t$
within accuracy $\epsilon$ turns out to be proportional
to $t^2/\epsilon$.  The elementary time increments
consist of the simulation of a local interaction for a small
time step $\Delta t=t/m$. Since $m\propto t^2$
the time steps scale as $\Delta t\propto 1/t$.  If time increments
can be implemented on a quantum computer with a native
gate with a gate time proportional to the simulated time
(`scaled gate'), $t_g\propto\Delta t \propto 1/t$, then
the overall simulation time $T$ scales as $T=m t_g \propto t$,
i.e., the simulation time $T$ is proportional to the simulated time $t$.
In many cases, scaled gates may not be available, but
it turns out that the method of higher-order Trotterization can
approach computation times $\propto t^{1+\eta}/\epsilon^\eta$
where  $\eta$  can be made arbitrarily small
\cite{Raeisi2012,Childs2018}. 
The concept of scaled gates is also related to the analog blocks in digital-analog quantum simulations \cite{Lamata2018}.
Similar estimates can be
made for the general class of sparse Hamiltonians \cite{Berry2007}.
It is also known that there is no sub-linear scaling of $T$ with the
simulated time $t$,
a restriction known as ``no fast-forwarding theorem''
\cite{Berry2007}.
\begin{figure}[b]
	\centering
	\includegraphics[width=0.9\columnwidth]{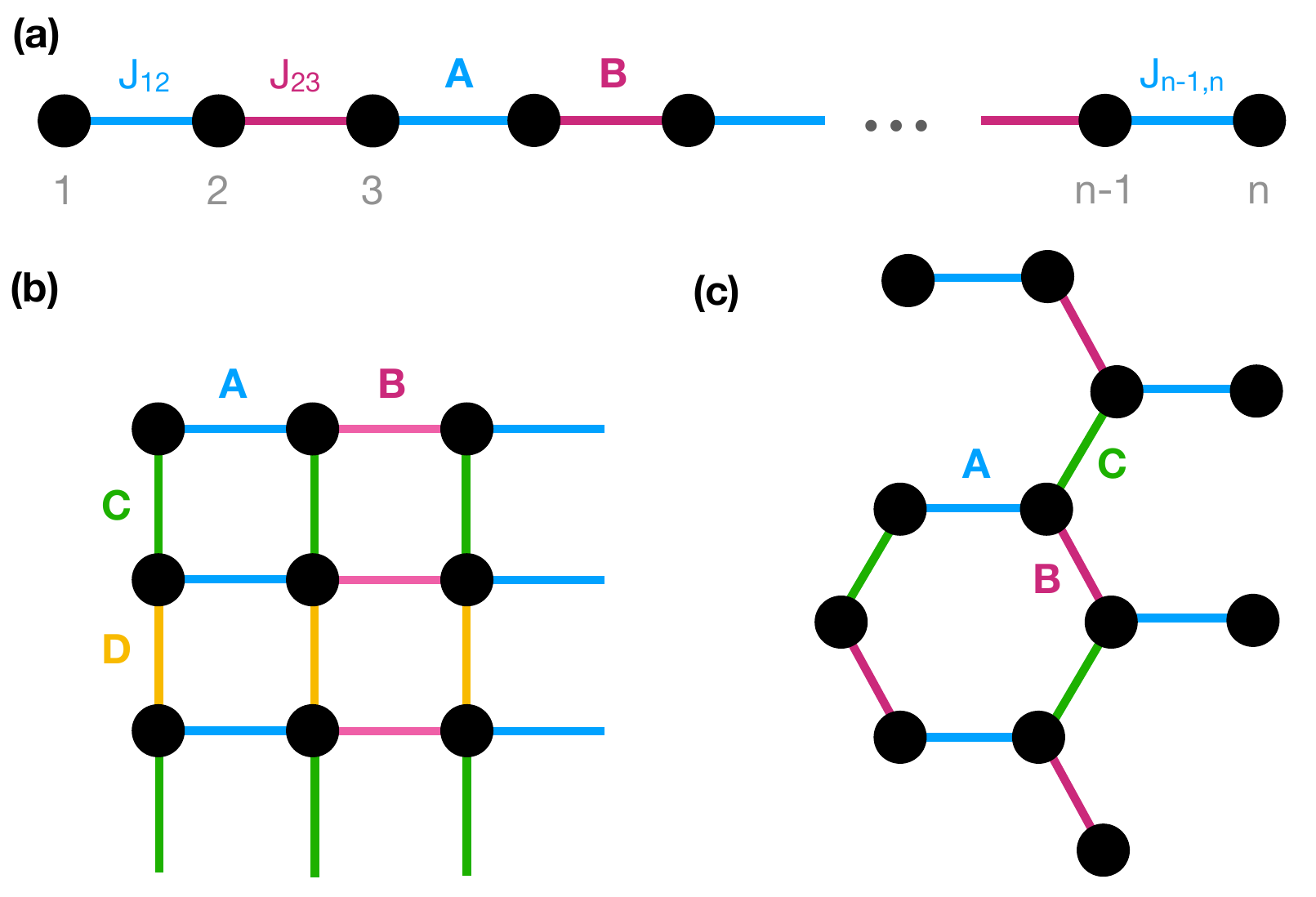}
        \caption{\textbf{Lattice spin system} with local interactions,
          in one (a) and two (b,c)
          spatial dimensions. A spin system can be described as a
          graph with vertices (edges) representing sites $i$ 
          (non-zero couplings $J_{ij}\neq 0$).
          In a regular lattice, the coordination
          number and the chromatic index $K$ are equal.
          The coordination number counts the number of
          nearest neighbors,
          while the chromatic index of a graph is the number of colors needed to
          color all edges without two same-colored edges meeting at any
          vertex.
          (a) One-dimensional lattice, $K=2$,
          (b) square lattice, $K=4$,
          (b) hexagonal lattice, $K=3$.}
    \label{fig:lattices}
  \end{figure}
  
Another question relates to the scaling of the simulation time $T$
with the size of the simulated system, e.g., measured in the number
of qubits $n$ required to store its quantum state.
Raeisi \textit{et al.} \cite{Raeisi2012} find that for $k$-local Hamiltonians $H=\sum_{j}
H_j$ where each $H_j$ acts on at most $k$ qubits, $T$ asymptotically
scales as $n^{(2+\eta)k-1} t^{1+\eta}/\epsilon^\eta$, again with
arbitrarily small $\eta$. If the number of qubits with which each qubit can
interact is bounded by a constant, the model is called
physically $k$-local, and $T$ is found to scale as $n^{1+\eta} t^{1+\eta}/\epsilon^\eta$,
i.e., nearly linear in both the simulated time $t$ and the system size
$n$ \cite{Raeisi2012}.

A useful observation is that for interacting sites described by a
regular graph, and in particular for lattices (such as those shown in
Fig.~\ref{fig:lattices}), graph coloring
according to Vizing's theorem allows for the decomposition of
the Hamiltonian into a number $K$ of commuting parts which does
not grow with the system size.  This allows for efficient
algorithms with simulation times asymptotically scaling as $t$,
and independent of $n$,
up to poly-logarithmic corrections in $nt/\epsilon$ \cite{Haah2018}.
In this paper, we study the Hamiltonian simulation
of spin models on a lattice that constitutes a digital quantum
simulation of physically 2-local Hamiltonians. We provide
explicit algorithms with the same asymptotic scaling.

The microscopic understanding of magnetic phenomena,
starting from ferromagnetism typically requires a quantum model \cite{Mattis2012}.
Magnetic behavior can be modeled by spin models
where spins are typically located on a lattice,
see Fig.~\ref{fig:lattices}.
The Heisenberg exchange interaction between the spins
is short-range and it is often an excellent approximation
to assume that only nearest neighbor spins are coupled.
Hamiltonian quantum simulation of such spin models
can provide useful insight into the time-dependent phenomena
of magnetic systems that is often hard to compute classically,
even in cases where the ground state of the system is
relatively easy to obtain.
Quantum circuits for  the digital quantum simulation of
disordered one-dimensional Heisenberg chains
have been developed and their scaling in $n$ and $t$
analyzed by Childs \textit{et al.} \cite{Childs2018},
where typically $t\propto n$
was chosen to simulate self-thermalization where
information needs to propagate through the entire system.

It is known that the Heisenberg interaction and other spin-spin
interactions--when combined
with local coupling of individual spins to an (effective) magnetic
field --generate a universal set of quantum gates for spin-1/2
qubits \cite{Loss1998}.   The Heisenberg interaction alone
can generate universal quantum computing on three-spin-1/2
exchange-only or decoherence-free subspace qubits
\cite{DiVincenzo2000}. Here, we will be concerned with the opposite
simulation direction, where a
universal quantum computer simulates a spin system.


\section{Spin models}

To begin, we define our general spin model, describing a  finite
number $n$ of spins $\boldsymbol{S}_i=(S_i^x, S_i^y, S_i^z)$ where $i=1,2,\ldots,n$
whose dynamics are described by the Hamiltonian
\begin{align}
  H  =\sum_{i< j} H_{ij}
      = \sum_{i< j} \sum_{\alpha\beta} J_{ij}^{\alpha\beta} S_i^\alpha S_j^\beta
  + \sum_{i\alpha} h_i^\alpha S_i^\alpha ,
  \label{eq:general}
\end{align}
where the spin operators fulfill the angular momentum commutation
rules, $\left[ S_i^\alpha,
  S_j^\beta\right]=i\delta_{ij}\sum_\gamma\epsilon_{\alpha\beta\gamma}S_i^\gamma$,
and where the interactions $ J_{ij}^{\alpha\beta}$ and external fields
$h_i^\alpha$ can be time dependent.
The length of the spin is arbitrary at this point but we will later
choose $S=1/2$ where each spin can be represented by one qubit.
We choose units in which $\hbar=1$ throughout this paper.
The site indices $i$ and $j$ run from $1$ to the number of sites
$n$, and the Cartesian coordinate indices $\alpha$, $\beta$, and $\gamma$ take
the values $x$, $y$, and $z$ for a three-dimensional spin.
The term $H_{ij}$ is defined such that it contains only spin operators $\boldsymbol{S}_i$ and
$\boldsymbol{S}_j$; such that we can, e.g., include the $h_i^\alpha$ terms in
$H_{i,i+1}$ for $i<n$ and in $H_{n-1,n}$ for $i=n$. Note that $\left[ H_{ij},
H_{i'j'}\right]=0$ for disjoint pairs $\left\{ i,j\right\} \cap
\left\{ i',j' \right\}=\emptyset$.
The connectivity graph defined by all nonzero $ J_{ij}$ tensors is
completely general at this point, but will be restricted below to
regular physical lattices that are constrained by locality and spatial
dimension.
The isotropic Heisenberg model represents an important special case
where $J_{ij}^{\alpha\beta}=\delta_{\alpha\beta}J_{ij}$ and thus
\begin{align}
H = \sum_{i<j}J_{ij} \boldsymbol{S}_i\cdot \boldsymbol{S}_j + \sum_i
  \boldsymbol{h}_i \cdot \boldsymbol{S}_i.
  \label{eq:Heisenberg}
\end{align}

The Hamiltonian generates the time evolution of the spin system in the
form of a time-ordered exponential
\begin{align}
  \psi(t)=\mathrm{T}\exp\left(-i\int_0^t  \!\! H(\tau)d\tau\right)\psi(0) =U(t)\psi(0),
  \label{eq:dynamics}
\end{align}
which can be approximated as a product of a finite number $m$ of
simple operator exponentials,
$U(t) \approx \prod_{p=0}^{m-1} \exp\left\{-i (t/m) H(pt/m)\right\}$
with the earliest times appearing on the right of the product.
%
If the Hamiltonian is time-independent, one has $U(t) = \exp( - i t H)$,
but this will not be assumed here.
In general, these exponentials cannot be
decomposed into factors $\exp(-itH_{ij})$ operating on spin pairs because various pairs
of terms in $H$ do not commute. To approximate $U(t)$, we
divide $H$ into a sum of $K$ non-commuting parts $H_k$ each of
which consists only of commuting terms,
\begin{align}
  H=\sum_{k=1}^K H_k, \quad\quad   H_k=\sum_{(i,j)\in P_k} H_{ij},
  \label{eq:partition}
\end{align}
with $\left[H_k,H_l\right]\neq 0$ for $k\neq l$.
Here, the sets $P_k$ are chosen such that they do not contain any
common spins,  and thus
$\left[H_{ij},H_{nm}\right]= 0$ for $(i,j),(n,m)\in P_k$, and thus
\begin{align}
\exp(-i \tau H_k) = \prod_{(i,j)\in P_k} \exp(-i \tau H_{ij})
  =\prod_{(i,j)\in P_k}  U_{ij}.
\end{align}

The problem of finding a minimal number $K$ of sets $P_k$ with this
property is equivalent to the edge coloring problem of the
graph $G$ consisting of $n$ vertices (one for each spin) and
an edge for each non-zero $J_{ij}$.  In graph theory, the minimal $K$ is
referred to as the chromatic index (Fig.~\ref{fig:lattices}).  Vizing's theorem \cite{Raeisi2012} states
that for a graph $G$ of degree $\deg(G)$, one has $\deg(G)\le K\le
\deg(G)+1$.  For bipartite lattices, $K=\deg(G)$.
Here, the degree $\deg(G)$ of $G$ is  defined as the maximum number
of edges containing the same vertex, i.e., the maximum number
of spins coupled to one and the same spin.
In the case of all-to-all coupling where all  $n(n-1)/2$ possible couplings between spins are assumed
to be nonzero, the graph describing the spin model
is the complete graph with degree $n-1$
and  $K=n-1$ if $K$ is even and $K=n$ if $n$ is odd.  In both cases, we find $K=O(n)$.

In the following, we will study spin models on a regular lattice
with local interactions, with each spin (in the bulk) being coupled to
at most its $\deg(G)=z$ adjacent spins where  the coordination number $z$ of the
lattice is independent of $n$.  Examples of lattice spin models in one and two dimensions
are shown in Fig.~\ref{fig:lattices}.
In this case $K=z$ and we will use the notation $K$ for both the
coordination number and chromatic index.
\begin{figure}
       \centering
	\includegraphics[width=0.8\columnwidth]{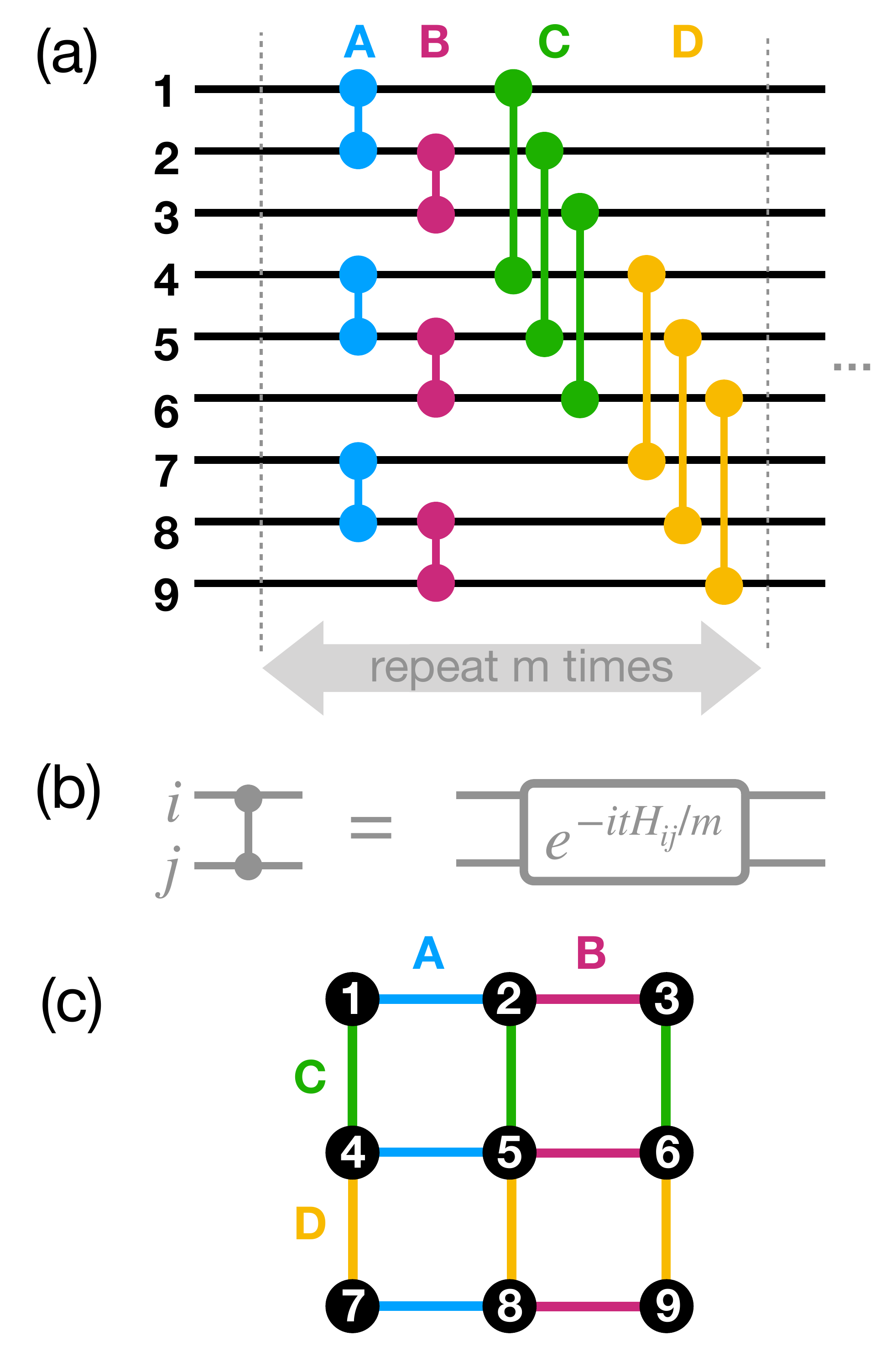}
        \caption{\textbf{Quantum circuit for the quantum simulation of
            a spin system.} (a) Quantum circuit for the Hamiltonian
          simulation of the lattice spin system shown in (c) with $K=4$
          within the first-order Trotterization. Colors and labels A,
          B, C, and D represent
          corresponding edges in the lattice.  The depicted sequence
          is repeated $m$ times where $m\sim K^2 t^2nJ^2$.
          (b) Elementary time increment simulating the 
          interaction between spins $i$ and $j$ during the time
          $t/m$.}
      \label{fig:circuit_sim}
    \end{figure}
    %


\section{First-order Trotter-Suzuki expansion}
    
\begin{figure*}
       \centering
	\includegraphics[width=\textwidth]{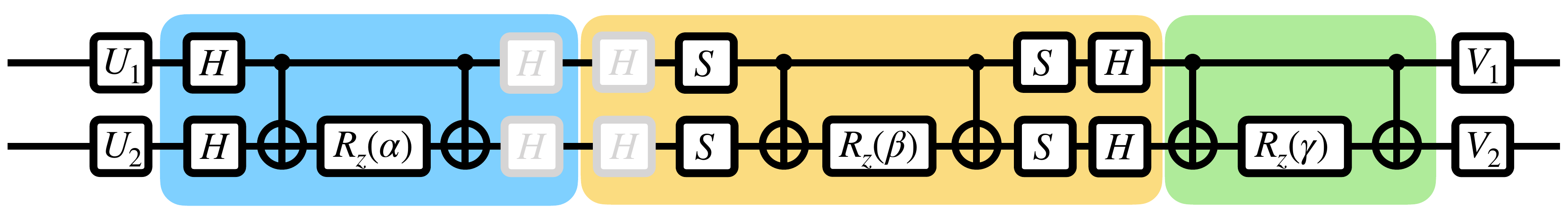}
        \caption{\textbf{Quantum circuit for a general two-spin
            interaction.}
          $U= V_1\otimes V_2\exp(-i\alpha S_{1}^x S_{2}^x -i\beta S_{1}^y S_{2}^y-i\gamma S_{1}^z S_{2}^z) U_1\otimes U_2$
          using controlled-NOT (CNOT) and single-qubit gates.
          The blue, yellow, and green shaded sections implement the
          commuting operations
          $\exp(-i\alpha S_{1}^x S_{2}^x)$, $\exp(-i\beta S_{1}^y
          S_{2}^y)$, and $\exp(-i\gamma S_{1}^z S_{2}^z)$,
          respectively.
       The greyed-out Hadamard gates can be omitted as they cancel
       each other.}
      \label{fig:general_circuit}
    \end{figure*}
    
The non-commutativity of the $H_k$ can be dealt with using the
Trotter-Suzuki expansion.
The first-order Trotter-Suzuki formula \cite{Trotter1959,Suzuki1977} allows for a digital quantum simulation
using alternating simulations of the non-commuting parts $H_k$
of $H=\sum_k H_k$,
\begin{align}
  e^{-it \sum_k H_k }
=\lim_{m\rightarrow\infty }
  \left( \prod_{k=1}^K e^{-i t H_k/m}\right)^m
  \!\!\!=\!\!\lim_{m\rightarrow\infty } S^{(1)}(t,m)^m.
  \end{align}
A practical quantum simulation will use only a finite number $m$ of
interactions and will thus incur an error 
\begin{align}
  \Delta_K^{(1)}(t,m)
 &=\left\lVert  e^{-it \sum_k H_k }-\left(
                        \prod_{k=1}^K e^{-i t H_k/m}\right)^m
                        \right\rVert \nonumber\\
 &= \frac{t^2}{2m}\left\lVert \sum_{k<l}\left[H_k,H_l \right]   \right\rVert
                  +O\left(\left(\frac{t}{m}\right)^3\right),
\end{align}
where the higher-order contributions can be more precisely
written as an exponential \cite{Childs2021}.
Using the upper bound
$\left\lVert \sum_{k<l}\left[H_k,H_l \right]   \right\rVert
\le \sum_{k<l} \left\lVert \left[H_k,H_l \right]   \right\rVert
\le \frac{K(K-1)}{2} \max_{k<l} \left\lVert \left[H_k,H_l \right]   \right\rVert$
and evaluating the commutators for the spin model
Eqs.~\eqref{eq:general}
and \eqref{eq:partition},
$\left[H_k,H_l \right]=i\sum_{i} J_{i i_k} J_{i
  i_l}\sum_{\alpha\beta\gamma}\epsilon_{\alpha\beta\gamma}S_i^\alpha
S_{i_k}^\beta S_{i_l}^\gamma$,
where $i_k$ is the unique site such that $(i,i_k)\in P_k$,
we find 
$\left\lVert\left[H_k,H_l \right]\right\rVert \le \frac{6}{8}n J^2$,
and
\begin{align}
  \Delta_K^{(1)}(t,m)\le \frac{3}{4}\frac{t^2}{2m}  \frac{K(K-1)}{2} n J^2,
\end{align}
where $J$ denotes an upper bound on the interaction strengths,
$J_{ij}\le J$.
In order to suppress the error below $\epsilon$,
such that $\Delta_K^{(1)}(t,m)\le\epsilon$, it is thus sufficient to
choose a sufficiently fine discretization of time, such that
\begin{align}
  m \ge \frac{3}{16}K(K-1)\frac{t^2}{\epsilon} n J^2.
\end{align}
The required number of elementary spin-spin coupling operations $U_{ij}=\exp(-i\tau
H_{ij})$ within the first-order Trotter-Suzuki expansion can then be given as
\begin{align}
N^{(1)}  = m\frac{n K}{2} =  \frac{3}{32}K^2(K-1)\frac{t^2}{\epsilon}
  n^2 J^2.
  \label{eq:gatecount1}
\end{align}
A quantum circuit realizing the digital quantum simulation of a spin
system is shown in Fig.~\ref{fig:circuit_sim}.
The circuit size (gate count) is proportional to $N^{(1)}$;  as we
show below, the CNOT count for general spin-spin interactions
amounts to $6N^{(1)}$, which is reduced to $3N^{(1)}$
in the case of Heisenberg interactions.
To quantify the circuit depth (simulation time) we observe
that the elementary spin-spin coupling operations $U_{ij}=\exp(-i\tau
H_{ij})$ inside each
$H_k$ can be executed in parallel (Fig.~\ref{fig:circuit_sim}a).
Therefore, we find for the simulation time,
\begin{align}
T^{(1)} =  m K t_g= \frac{3}{16} K^2(K-1)\frac{t^2}{\epsilon}
  n J^2 t_g,
  \label{eq:runtime1}
\end{align}
where $t_g$ is the maximum time required to execute $U_{ij}=\exp(-i\tau
H_{ij})$ with the simulated time increment $\tau=t/m$.
The quantum circuit for $U_{ij}$ may contain `fixed' quantum gates that
require a gate time that is independent of $\tau$, others may be
`scaled', i.e., require a gate time proportional to $\tau$ (see also the concept of digital-analog simulation \cite{Lamata2018}).
In the case of a digital quantum simulation, there are some fixed
gates, such as CNOT, and thus  $t_g=t_\infty+ s t/m$
with $t_\infty>0$, $s\ge 0$. Assuming that some of the used
quantum gates (e.g., CNOT, H, etc.) have a fixed gate time, and
noting that for large $m$ the contribution of scaled gates to the
gate time is small and can be bounded by a constant, we set
$s=0$ and $t_\infty>0$, and thus $t_g=t_\infty={\rm const.}$

The results for the circuit depth and simulation run time,
Eqs.~\eqref{eq:gatecount1} and \eqref{eq:runtime1}, do not
achieve the best possible scaling in $n$ and $t$.
In the following two sections, we discuss two possibilities
to further improve the scaling.  On the one hand, one can
resort to higher-order Trotter-Suzuki expansions within
digital quantum simulation. On the other hand, if scaled
gates are available, one can proceed without the use of
higher-order  Trotter-Suzuki expansions.


\section{Higher-order Trotter-Suzuki expansion}

The second-order and higher-order Trotter-Suzuki formulas
\cite{Suzuki1991} can be written as
    \begin{align}
S^{(2)}(t,m)= &   \prod_{k=1}^K e^{-i t H_k/2m}\prod_{k=K}^1
                 e^{-i t H_k/2m},\\
      S^{(2q)}(t,m)= &  \left( S_{2p-2}(p_q t,m)\right)^2
                        S_{2p-2}((1-4p_q) t,m)\nonumber\\
                        &  \times \left( S_{2p-2}(p_q t,m)\right)^2,
    \end{align}
    with $p_q=(4-4^{1/(2q-1)})^{-1}$ for $q>1$.
    As in the first-order case, $\lim_{m\rightarrow
      \infty}S^{(2q)}=e^{-it\sum_k H_k}$, for all $q\ge 1$, but the
    convergence becomes faster for higher orders of the
    Trotter-Suzuki formula.  This leads to improved error
    bounds which can be found in Refs.~\cite{Wiebe2010, Barthel2020,Childs2021},
        \begin{align}
\Delta_K^{(2q)}(t,m)
 &=\left\lVert  e^{-it \sum_k H_k }-\left( S^{(2q)}(t,m)\right)^m
   \right\rVert\\
          &=          c_1 \frac{t^{2q+1}}{m^{2q}} \sum_{k_1,\ldots, k_{2q+1}}^K \left\lVert
          \left[H_{k_{2q+1}},\ldots \left[H_{k_2}, H_{k_1}\right]\right]\right\Vert ,
        \end{align}
where $c_1$ is a constant tat can depend on $q$ (similarly for all $c_i$ below).
        Evaluating the commutators for a local spin Hamiltonian, we find for
the error
                \begin{align}
\Delta_K^{(2q)}(t,m)
 & \le c_2 \frac{(Kt)^{2q+1}}{m^{2q}}n .
        \end{align}
        Keeping the error below $\epsilon$ then requires,
         \begin{align}
m \ge c_3 \frac{(K t)^{1+1/2q}}{\epsilon^{1/2q}} n^{1/2q},
         \end{align}
which leads to an interaction gate count of
         \begin{align}
N^{(2q)}  = c_4 m\frac{n K}{2} \gtrsim c_5 \frac{n^{1+1/2q} K^{2+1/2q} t^{1+1/2q}}{\epsilon^{1/2q}} .
\end{align}
Again assuming that the interaction gates simulating $H_{ij}$ inside
each  $H_k$ can be executed in parallel, we find for the simulation time,
\begin{align}
T^{(2q)} =  c_4 m K t_g\gtrsim c_6 \frac{K^{2+1/2q}
  t^{1+1/2q}}{\epsilon^{1/2q}} n^{1/2q} t_g.
\end{align}
The fact that both the number of interaction gates $N^{(2q)}$ and
the simulation time $T^{(2q)}$ can in principle be made to scale
arbitrarily close to linearly in the simulated time $t$, has been
pointed out in Ref.~\cite{Berry2007}.  Also the vanishing influence of
the target accuracy $\epsilon$ on the simulation time with increasing order has been recognized.  The fact that the exponent of the
problem size $n$ can also be suppressed is specific to 
constructions where the interaction gates for commuting
interactions \cite{Haah2018}, as illustrated in Fig.~\ref{fig:circuit_sim}.


\section{Implementation}
\begin{figure}[b]
       \centering
	\includegraphics[width=\columnwidth]{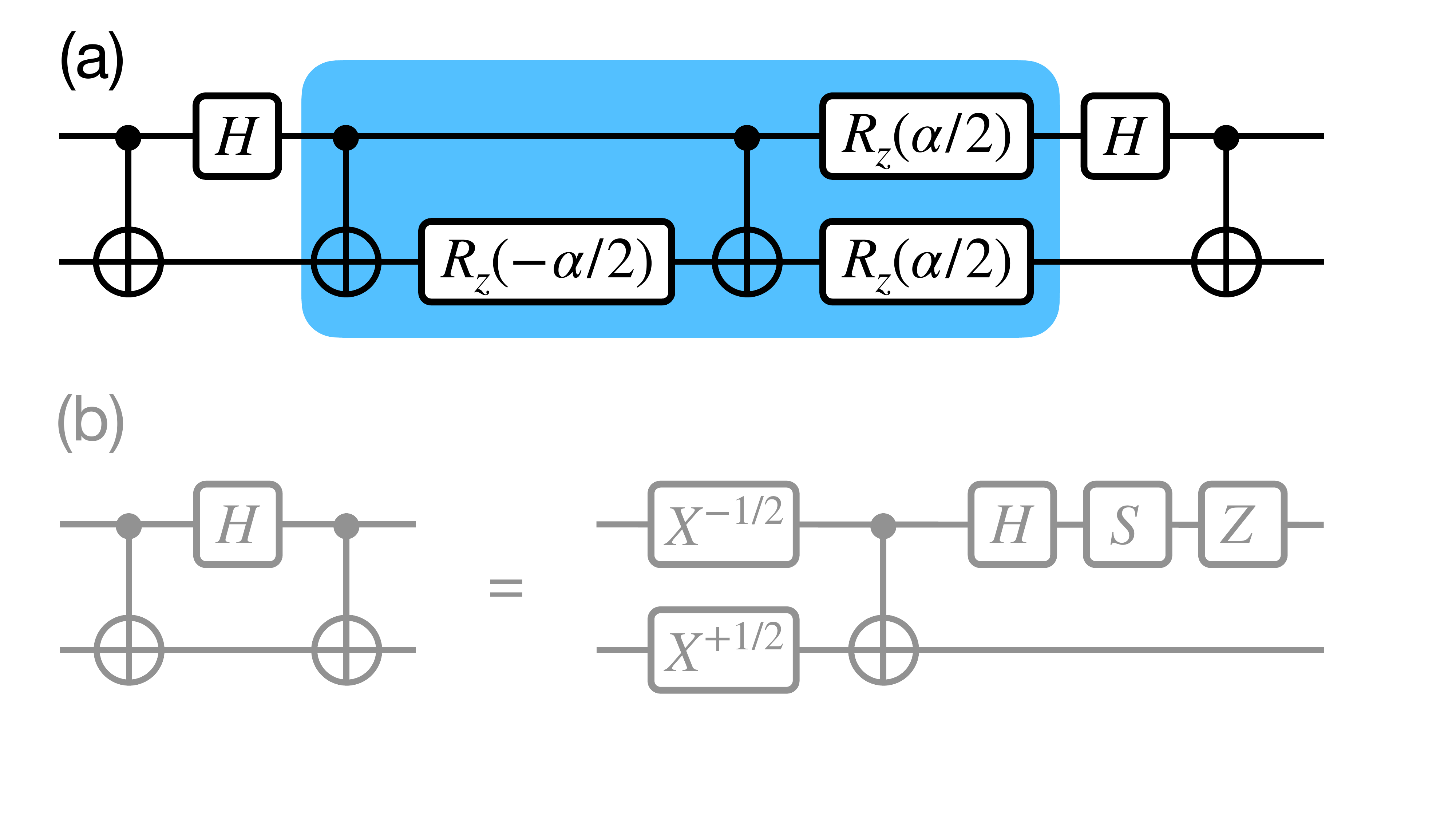}
        \caption{\textbf{Quantum circuit for the Heisenberg interaction.}
          (a) Implementation of $\exp(-i\alpha
          \mathbf{S}_1\cdot\mathbf{S}_2)=\mathrm{SWAP}^{\alpha/\pi}$
          using controlled-NOT (CNOT) and single-qubit gates.
          The blue-shaded section implements a
          controlled-phase gate CPHASE($\alpha$).
        (b) Replacement for first three gates in (a) that eliminates
        one CNOT gate.}
      \label{fig:Heisenberg_circuit}
    \end{figure}

To realize elementary interaction gates $U_{ij}$, we use the
decomposition \cite{Zhang2003}
$U=\exp(-i\tau H_{ij})
=V_1\otimes V_2\exp(-i\alpha S_{1}^x S_{2}^x -i\beta S_{1}^y
S_{2}^y-i\gamma S_{1}^z S_{2}^z) U_1\otimes U_2$,
where
$H_{ij}=\sum_{\alpha\beta} J_{ij}^{\alpha\beta} S_i^\alpha S_j^\beta
+ \sum_{\alpha} (h_i^\alpha S_i^\alpha+h_j^\alpha S_j^\alpha) $,
and where $U_{1,2}$ and $V_{1,2}$ are one-qubit gates.
This unitary can be assembled using elementary gates with the
circuit shown in Fig.~\ref{fig:general_circuit}, requiring six CNOT gates.
The special case of Heisenberg interactions allows for a
simpler circuit with only three CNOT gates, as shown
in Fig.~\ref{fig:Heisenberg_circuit}.

    The gate implementations comprise both fixed-length gates
    such as Hadamard and CNOT, and the scaled gate $R_z(\alpha)$
    where $\alpha \propto \tau$ for an interval $\tau\propto 1/m$ of simulated
    time.  With increasing number $m$ of time intervals, the
    fixed-length gates will dominate the execution time of such
    interaction gate constructions.

    Depending on the quantum hardware, the interaction gates $U=\exp(-i\tau H_{ij})$
    for some $H_{ij}$ may be native, i.e., directly implementable
    in time proportional to $\tau$, similar to the analog blocks in digital-analog simulation \cite{Lamata2018}.  We call this implementation a
    {\em scaled gate}.  If all required interaction gates are
    available as scaled gates, then $t_g = s t/m$ with $s$ a constant
    and $t_\infty=0$, and Eq.~\eqref{eq:runtime1} turns into
    \begin{align}
      T^{(1)} \le K s t,
    \end{align}
    with $K$ and $s$ constants describing the degree of the lattice
    graph and the ratio between simulation time and simulated time
    for the slowest scaled gate.  In this case, we obtain a simulation
    time linear in the simulated time already using the first-order Trotter
    expansion.  The higher-order Trotter expansions do not provide any
    improvement in this case.
    The exclusive use of scaled gates renders the simulation time
    $T$ independent of the number $m$ of discrete time steps.
    Therefore, at least for time-independent problems, one can choose
    $m=1$ which corresponds to a direct analog quantum simulation.


    \section{Conclusions}
   We have shown explicit circuit constructions that realize the
   resource estimates for first-order and higher-order Trotter-Suzuki
   product formulas. 
We conclude with an open question.  Given a set of available native one- and
two-qubit scaled gates, which other scaled gates can be efficiently constructed
with this set?  Do universal sets of scaled gates exist that allow for
the synthesis of arbitrary scaled gates $U_{ij}$?  Answering
these questions may give further insight into which simulation tasks
can be performed even more efficiently than shown here.


\bibliography{SpinSim}

\end{document}